\documentclass[twocolumn,showpacs,preprintnumbers,amsmath,amssymb]{revtex4}

\def\ube13{UBe$\rm_{13}$}

\def\bi2212{Bi$\rm_2$Sr$\rm_2$CaCu$\rm_2$O$\rm_8$}
\def\ybi2212{Bi$\rm_2$Sr$\rm_2$YCu$\rm_2$O$\rm_8$}
\def\ycabi2212{Bi$\rm_2$Sr$\rm_2$Ca$\rm_{1-x}$Y$\rm_x$Cu$\rm_2$O$\rm_{8+\delta}$}

\def\y65cabi2212{Bi$\rm_2$Sr$\rm_2$Ca$\rm_{0.35}$Y$\rm_{0.65}$Cu$\rm_2$O$\rm_{8+\delta}$}

\def\Co{CeCoIn$_5$}
\def\Rh{CeRhIn$_5$}

\usepackage{graphicx}
\usepackage{dcolumn}
\usepackage{bm}

\begin{document}

\preprint{submitted to Physical Review Letters \hfill LA-UR-03-0364}

\title{Avoided Antiferromagnetic Order and Quantum Critical Point in CeCoIn$_5$}

\author{A. Bianchi, R. Movshovich, I. Vekhter, P. G. Pagliuso, and J. L. Sarrao}
\affiliation{%
Los Alamos National Laboratory, Los Alamos, New Mexico 87545
}%


\date{\today}

\begin{abstract}

We measured specific heat and resistivity of heavy fermion \Co\ between the superconducting
critical field $H_{c2} = 5 T$ and 9 T, with field in the [001] direction, and at temperatures down
to 50mK. At 5T the data show Non Fermi Liquid behavior down to the lowest temperatures. At field
above 8T the data exhibit crossover from the Fermi liquid to a Non Fermi Liquid behavior. We
analyzed the scaling properties of the specific heat, and compared both resistivity and the
specific heat with the predictions of a spin-fluctuation theory. Our analysis leads us to suggest
that the NFL behavior is due to incipient antiferromagnetism (AF) in \Co\, with the quantum
critical point in the vicinity of the $H_{c2}$. Below $H_{c2}$ the AF phase which competes with the
paramagnetic ground state is superseded by the superconducting transition.

\end{abstract}

\pacs{74.70.Tx, 71.27.+a, 74.25.Fy, 75.40.Cx}
\maketitle

When the symmetry of the ground state of a system changes as a function of an external or internal
parameter, the system is said to undergo a quantum phase transition. If, in addition, this
transition is second order, the system has a Quantum Critical Point (QCP) at the critical value of
the parameter. The competition between the nearly degenerate ground states determines the behavior
of the system over a range of temperatures and tuning parameter values in the vicinity of QCP. In
this region of the phase diagram the properties of the system differ from those on either side of
the transition, and often exhibit unusual dependence on the temperature and the tuning parameter.
This has made quantum critical phenomena a subject of intense current interest.

Study of quantum critical points in heavy fermion systems has been a focus of particular attention
(for a recent review see Ref.~\onlinecite{stewart:rmp-2001}). In these materials the competition
typically takes place between a paramagnetic and a magnetically ordered ground states. The
unconventional behavior near QCP is manifested in the deviation of the temperature dependence of
measured properties from those of metals described by the Landau Fermi Liquid (FL) theory. In that
theory the electronic specific heat is linear in temperature, $C(T)=\gamma T$, and the resistivity
increases quadratically from a residual value, $\rho=\rho_0 + AT^2$. In systems tuned to QCP the
Sommerfeld coefficient, $\gamma(T) = C/T$, commonly diverges as the temperature goes to zero, and
has been variously argued to behave as either $\log T$ or $T^\alpha$, with $\alpha <0$. Resistivity
with an exponent less than two is also ubiquitous in these compounds.

Tuning the system through a QCP  can be accomplished experimentally by varying sample's
composition~\cite{lohneysen:prl-94,maple:jltp-95}, applying pressure~\cite{mathur:nature-98}, or
applying magnetic field~\cite{grigera:nature-01}. In non-stoichiometric compounds the Kondo
disorder, where a range of Kondo temperatures $T_K$ appears due to different environments of the
f-electron ions, is an important mechanism leading to a Non Fermi Liquid (NFL)
behavior~\cite{bernal:prl-95,maclaughlin:jpcm-96}. In these compounds it is not easy to separate
this origin of NFL behavior from the consequences of the proximity to a QCP. Hence the
stoichiometric compounds receive great deal of attention in the field of quantum criticality.

One class of such materials are Ce-based compounds, which have an antiferromagnetic (AF) ground
state at ambient pressure. Hydrostatic pressure suppresses the magnetic ordering temperature $T_N$
to zero at a critical pressure of the QCP. Such approach was used successfully for
CeCu$_2$Ge$_2$~\cite{jaccard:pla-92} CeRh$_2$Si$_2$~\cite{movshovich:prb-96}, CePd$_2$Si$_2$,
CeIn$_3$~\cite{mathur:nature-98}, and some other compounds. Alternatively, AF order can be
suppressed by an applied magnetic field. When this was done in YbRh$_2$Si$_2$, an NFL behavior was
revealed again on the paramagnetic side of the QCP ~\cite{gegenwart:prl-02}.

In this Letter we present the results of the specific heat and resistivity measurements on
stoichiometric \Co, which show that this compound has a QCP with magnetic field as a tuning
parameter. We show that the quantum critical behavior is most likely due to the proximity of an
antiferromagnetic state. This case is particularly interesting because \Co\ is an ambient pressure
superconductor. We argue that magnetic order which competes with the paramagnetic state is
superseded by the superconducting state.

\Co\ is a tetragonal, quasi-2D compound, with layers of CeIn$_3$ separated by layers of CoIn$_2$.
It is an ambient pressure heavy fermion superconductor~\cite{petrovic:jpcm-01} with $T_c = 2.3$ K,
the highest value for this class of compounds. Superconductivity in \Co\ is unconventional, with
lines of nodes in the energy gap, as demonstrated by specific heat and thermal
conductivity~\cite{movshovich:prl-01} and NQR~\cite{kohori:prb-01} measurements. Pauli limiting
analysis~\cite{movshovich:sces-01} and thermal conductivity modulations in magnetic
field~\cite{izawa:prl-01} indicate that \Co\ is a singlet, $d_{x^2-y^2}$ superconductor. Recently
it was shown that the superconducting transition in \Co\ in magnetic fields close to the upper
critical field for [001] direction, $H_{c2} = 4.95 T$, is first order below $T = 0.7$
K~\cite{izawa:prl-01,bianchi:prl-02}. NFL behavior of \Co\ at the field of 5 T persists over a
large region of temperature~\cite{kim:prb-01}, with $\gamma \propto -\log(T)$ between 0.4 K and 8
K.

\begin{figure}
\includegraphics[width=3in]{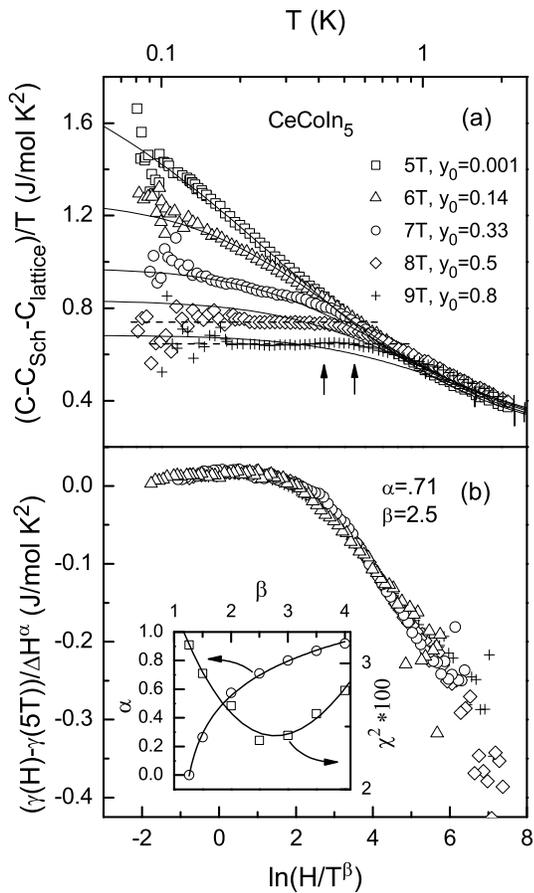}
\caption{(a) Sommerfeld coefficient, $\gamma(T)=C(T)/T$,of \Co\ in magnetic fields $H \parallel
$[001]. Dashed lines for 8T and 9T emphasize the FL behavior with constant $\gamma$. Left (right)
arrow indicate $T_{FL}^C$ for 8 T (9 T). Solid lines are fits to the SCR spin-fluctuation model for
each field with the corresponding values of $y_0$, see text for details. (b) scaling analysis of
the data in (a) for $\alpha = 0.71$ and $\beta = 2.5$. Inset: For different values of $\beta$ we
plot values of $\alpha$ which minimize $\chi^2$ for a given $\beta$, and $\chi^2$ for these
$\alpha$ and $\beta$.}

\label{hc-in-high-field}
\end{figure}

Here we concentrate on the low temperature region down to 50 mK and magnetic fields between $H_{c2}
= 4.95$ T and 9 T. Figure ~\ref{hc-in-high-field}(a) shows Sommerfeld coefficient of \Co\ versus
temperature for several magnetic fields between 5 T and 9 T. Large low temperature Schottky anomaly
tails due to the quadrupolar and magnetic spin splitting of In and Co nuclei were
subtracted~\cite{movshovich:prl-01}. At high magnetic field and low temperature Schottky term
accounts for most of the specific heat measured, and its subtraction results in an increased
scatter of the data points at low temperature. The data for 5 T follow logarithmic temperature
behavior below 1 K down to the lowest temperature studied. As the magnetic field is increased above
5 T, the low temperature data starts to deviate from the logarithmic behavior at ever higher
temperature. Clear FL regime of $\gamma =$ const. at low temperature is recovered at 8 Tesla. Both
8 T and  9 T  data exhibit sharp crossovers between the NFL (logarithmic) and the FL regimes at a
temperature $T_{FL}^C$. For fields below 8 T $\gamma$ continues to rise with decreasing temperature
and does not show saturation within the temperature range shown. This behavior is different from
that of YbRh$_2$Si$_2$, where $\gamma$ saturates below some temperature $T^*(H)$ to a constant
value for all $H \neq H_c$, with a rather sharp break from the $-\log(T)$ behavior for $T >
T^*$~\cite{trovarelli:prl-00}.

The logarithmic divergence of $\gamma(T, H)$ at 5T and $\gamma$'s increase followed by a crossover
to the FL behavior at low $T$ for higher fields is indicative of a field-tuned QCP near 5T. This
observation is strongly supported by the scaling properties of $\gamma$. For a phase transition
with $T_c=0$, the temperature alone sets the energy scale of fluctuations. Scaling  of dynamical
properties with energy, $E$, as $E/T$ is taken as good evidence of quantum criticality
~\cite{varma:prl-89,sachdev:qcp-book}.

In our case the tuning parameter is the applied magnetic field. Hence the energy scale of
fluctuations away from the QCP depends on $|H-H_c|$. Therefore the entropy, and hence the specific
heat, should scale as $\gamma(H)-\gamma(H_c) \propto f((H-H_c)/T^\beta)$. Experimentally, at fields
sufficiently away from the QCP, we achieve the FL regime at low $T$. In that regime $\gamma$
depends on the field but is $T$-independent. Consequently, the scaling relation has to take the
form $\gamma(H)-\gamma(H_c) \propto (H-H_c)^\alpha f((H-H_c)/T^\beta)$. The form of scaling is
similar to that obtained for U$_{0.2}$Y$_{0.8}$Pd$_3$~\cite{andraka:prl-91},
CeCu$_{5.8}$Ag$_{0.8}$~\cite{heuser:prb-98}, and for YbRh$_2$Si$_2$~\cite{trovarelli:prl-00}.

We find that with the choice of $H_c = 5$ T, the best scaling is achieved for $\alpha = 0.71$ and
$\beta = 2.5$; it is shown in Fig. 1(b). It is a remarkably good scaling, spanning both FL and NFL
regimes, with all four data sets for different fields overlapping each other in the entire
experimental temperature range. This scaling is a strong indication that the behavior of CeCoIn$_5$
in the part of the phase space explored in these experiments is governed by the QCP very close to
$H_c = 5$ T. The high power of temperature in the argument of the scaling function (the scaling
dimension of the magnetic field) suggests that the field is very efficient in suppressing the
critical fluctuations.

Importantly, scaling implies that the behavior observed here is associated with a second order
transition. Therefore it is unlikely that first order transition from the superconducting to normal
state above 4.7 T~\cite{bianchi:prl-02} controls the properties in the parameter range investigated
here, and we need to consider alternative competing orders.

\begin{figure}
\includegraphics[width=3in]{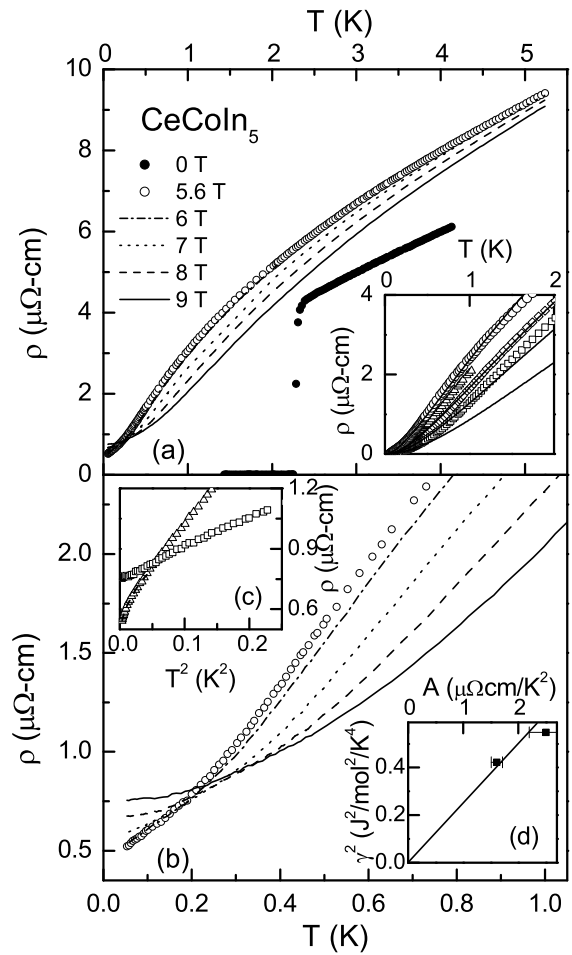}
\caption{(a) Resistivity of \Co\ in magnetic field $H
\parallel$ [001] above the critical field $H_{c2} = 4.95$
T. Inset: Fits to $\rho(T)$ below 2K from the spin-fluctuation theory as detailed in the text.
($\circ$) 6 T, ($\triangle$) 7 T, ($\diamond$) 8 T, ($\square$) 9 T. Solid lines: $y_0 =$ 0.14,
0.33, 0.5, and 0.8 from top to bottom. (b) Low temperature region of data in (a). Inset (d):
$\gamma^2$ vs. A. Solid line is a linear fit to the data points and the origin. Inset (c)
Resistance for 6 T ($\triangle$) and 9 T ($\square$) plotted vs. $T^2$. } \label{res-in-high-field}
\end{figure}

The crossover from the NFL to FL regimes is also clear from resistivity measurements in the same
magnetic field range. Fig.~\ref{res-in-high-field}(a) shows resistivity, $\rho$, of \Co\ below 5 K
for fields between 5.6 T and 9 T, as well as the zero-field data. The data at 5.6 T and above is
not affected by the proximity to the superconducting transition, as seen from the magnetoresistance
measurements of \Co\ at 100 mK.~\cite{movshovich:unpublished} To focus on the low temperature
behavior, we show the data below 1 K in Fig.~\ref{res-in-high-field}(b). The data for 8 T below 300
mK and for 9 T  below 500 mK display the FL-like $\Delta\rho\equiv\rho-\rho(T=0)=A T^2$ behavior,
consistent with the linear in $T$ specific heat data for these fields. In contrast, for fields of
5.6 T and 6 T we found $\Delta\rho\propto T$ below 200 mK, and the data for 7 T show intermediate
behavior, the best fit being the sum of linear and quadratic terms in this temperature range. At
these fields the linear in $T$, rather than quadratic, dependence of the resistivity is consistent
with the NFL behavior of the specific heat. Qualitative difference in the temperature behavior of
$\rho(T)$ at 6T and at 9T in the low temperature range is emphasized in the upper inset in
Fig.~\ref{res-in-high-field}(b), where resistivity is plotted versus $T^2$ for two fields, 9 T and
6 T.  The 9 T data are linear in $T^2$ over the entire temperature range shown, while the 6 T data
has a pronounced negative curvature.

To further analyze the FL region of the phase diagram of \Co\, in the inset (d) of
Fig.~\ref{res-in-high-field}(b) we plot $\gamma^2$ versus $A$ for the 8 T and 9 T data. Within
error bars the two values lie on the line $\gamma^2 = .3 \times A$ $\rm J^2/ mol^2K^2\mu \Omega
cm$. The slope is about three times larger than the Kadowaki-Woods ratio ~\cite{kadowaki:ssc-86},
approximately obeyed by many heavy fermion compounds. This discrepancy is within the typical range
of the scatter of the data for other HF compounds, and confirms the FL ground state of \Co\ for 8 T
and 9 T.

Having established the existence of the QCP near $H_c\approx 5$ T, we proceed to identify the order
competing with the paramagnetic FL ground state. Since a closely related compound, CeRhIn$_5$, is
an ambient pressure antiferromagnet with the Neel temperature $T_N=3.8$ K, AF order is a natural
candidate. It is also consistent with the $d_{x^2-y^2}$ superconductivity.

While the complete theory of quantum criticality in itinerant antiferromagnets is still lacking,
the singular contribution of the critical AF spin fluctuations to the specific heat and scattering
has been considered in Ref.~\onlinecite{moriya:jpsj-95} in the framework of a self-consistent
renormalization (SCR) theory. This theory was used to analyze the specific heat of
CeCu$_{5.2}$Ag$_{0.8}$ near the QCP~\cite{stewart:rmp-2001}. The input parameter of the theory is
the distance from a QCP, $y_0$, ($y_0=0$ at the QCP), related to the inverse square of the magnetic
correlation length. The specific heat and the resistivity are functions of the reduced temperature,
$T/T_0$, where $2\pi^2 T_0$ is of the order of the exchange interaction. We use this theory to fit
our data.

Close to the QCP we obtained a good fit to both the specific heat data (shown in Fig. 1(a)), and
resistivity (shown in the inset of Fig. 2(a)) for the same parameter values. To estimate $T_0$, we
compared \Co\ with \Rh. Pressure of 16 kbars suppresses the AF state in \Rh, at which point the
superconducting state with $T_c = 2.1$ K emerges\cite{hegger:prl_00}. This lead Sidorov {\it et
al.} to suggest that physical properties of \Co\ are very close to those of \Rh\ at 16
kbars~\cite{sidorov:prl-02}. The $T_0 = 0.4$ K we chose is consistent with the exchange energy of
the order of a few Kelvin, expected from the $T_N = 3.8$ K for \Rh. In the fit we assumed an
additive field-independent contribution to $\gamma(T)$ of 0.2 J/mol K$^2$ from non-critical
fermions. For the 5 T data the theoretical fits are identical for any $y_0\leq 0.01$ over the
temperature range studied. Consequently, \Co\ is very close to a QCP at this field.

Since in our experiments the magnitude of the applied field (used to tune the system to the QCP) is
comparable to the exchange interaction, we expect that the changes in the spectrum of the spin
fluctuations and the quasiparticle scattering rate  lead to deviations from the predictions of the
SCR theory when the field is increased away from QCP. Indeed, at higher fields the discrepancy
between the fits and the resistance data  becomes more pronounced (especially at higher
temperature). However, the overall S-shape of the resistance curves agrees with the data. We
therefore conjecture that the QCP is due to the tendency towards the AF order.

\begin{figure}
\includegraphics[width=3in]{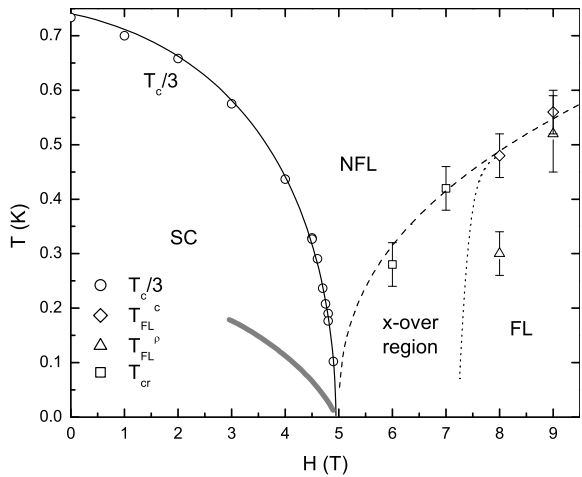}
\caption{Combined phase diagram of \Co. Superconducting-normal phase boundary (($\circ$) and solid
line) is from Ref.~\onlinecite{bianchi:prl-02}. $T_{cr}$ ($\square$) and the dashed line (guide to
eye) denote the upper boundary of the crossover region. Dotted line through $T_{FL}^{C}$
($\diamond$) and $T_{FL}^{\rho}$ ($\triangle$) schematically indicate the upper boundary of the FL
region. Region between the dashed and dotted lines is the crossover region. Thick grey line is the
hypothetical phase boundary of the AF state prevented by the superconducting transition.}
\label{phase-diagram}
\end{figure}

Finally, in Fig.~\ref{phase-diagram} we plot the phase diagram of \Co\ in the $T-H$ plane. The
normal-superconducting phase boundary was determined in
Refs.~\onlinecite{petrovic:jpcm-01,bianchi:prl-02}. At fields of 8T and 9T we see a clear
transition from the NFL to FL behavior in both specific heat ($T_{FL}^C$) and resistivity
($T_{FL}^\rho$). Below 8T, in the temperature range studied, the heat capacity and resistance
deviate from the critical behavior (5 T data), although the FL regime is not reached. For these
fields we define a temperature $T_{cr}$ below which $\gamma$ begins to deviate from $-\log(T)$
behavior, leading to a broad crossover region. If the FL behavior exists at these fields, it occurs
over a narrow range of very low temperatures.

In conclusion we find that, while there is no signature of the AF ordered phase, the behavior of
\Co\ above the superconducting upper critical field, $H_{c2}=4.95$T, at low temperatures is
controlled by a field-tuned antiferromagnetic quantum critical point. The existence of the QCP is
confirmed by scaling analysis of the specific data. Incipient AF state in \Co\ is strongly
suggested by the the fits to both the specific heat and the resistivity based on the
spin-fluctuation theory~\cite{moriya:jpsj-95}. The underlying physical picture is that the AF spin
fluctuations promote anti-alignment of the electron spins. Whether time-reversal breaking
(antiferromagnetism) or singlet formation with phase coherence (superconductivity) occurs first
depends on the details of the system. In \Co\ the superconductivity prevails, preventing the
magnetic order from developing. However, there are no critical fluctuations associated with the
(first order ~\cite{bianchi:prl-02}) superconducting transition at $H_{c2}$, so that the behavior
above $H_{c2}$ is controlled by the proximity to the AF critical point. Our results imply that the
transition temperature of the avoided antiferromagnetic order vanishes at fields very close to the
superconducting upper critical field. This concomitant suppression of the AF and SC orders is
unusual, and may reflect an important aspect of the underlying physics, presenting further
challenges to theory and experiment.

We thank J. D. Thompson, C. Varma, Q. Si, P. Coleman, and G. R. Stewart for stimulating
discussions. We are also grateful to  J. D. Thompson for help with selecting Indium-free samples.
Work at Los Alamos National Laboratory was performed under the auspices of the U.S. Department of
Energy.


\end{document}